# A configurable computer simulation model for reducing patient waiting time in oncology departments

R. R. Corsini[a], A. Costa[b], S. Fichera[b] and A. Pluchino[a,c]

[a]Department of Physics and Astronomy "E-Majorana", University of Catania, Italy;

[b]DICAR Department, University of Catania, Italy;

[c]Sezione INFN of Catania, Italy

**Abstract:** Nowadays, the increase in patient demand and the decline in resources are lengthening patient waiting times in many chemotherapy oncology departments. Therefore, enhancing healthcare services is necessary to reduce patient complaints. Reducing the patient waiting times in the oncology departments represents one of the main goals of healthcare manager. Simulation models are considered an effective tool for identifying potential ways to improve patient flow in oncology departments. This paper presents a new agent-based simulation model designed to be configurable and adaptable to the needs of oncology departments which have to interact with an external pharmacy. When external pharmacies are utilized, a courier service is needed to deliver the individual therapies from the pharmacy to the oncology department. An oncology department located in southern Italy was studied through the simulation model and different scenarios were compared with the aim of selecting the department configuration capable of reducing the patient waiting times.

## 1. Introduction

While tackling high healthcare costs and restricted budgets, oncology departments have faced new managerial challenges stemming from the need to satisfy the ever increasing patient demand. The worldwide ageing population has resulted in an increase in demand for healthcare services. Healthcare systems are able to increase the population's life expectancy and, consequently, the mean age of the population. At the same time, ageing involves a decrease in immune defences and an enlarged predisposition to illness (Candore et al., 2006; Troen, 2003). In the opinion of Vasto et al. (2009), "*the pro-* *inflammatory status of ageing might be one of the both convergent and divergent mechanisms which relate cancer to ageing*". Furthermore, epidemiological studies have shown causal associations between cancers and several other factors, such as lifestyle or diet (Katzke et al., 2015), tobacco exposure (Montesano & Hall, 2001) and air pollution (Martoni, 2018). Siegel et al. (2021) estimated 1,898,160 cases of cancer in the year 2021 in the United States. Due to this, a steady growth in demand for healthcare services within oncology departments has been seen. However, patient demand for care is not always met by adequate levels of services within oncology departments. The outcomes of this dichotomy are a higher workload for healthcare providers and a dramatic growth in patient waiting time. Oncology facilities manage large volumes of patients with limited resources (*e.g.*, pharmacists, nurses or treatment chairs) (Alvarado et al., 2018). Goldstein et al. (2008) stated that the balance between supply and demand of total annual oncology visits in 2005 would become unbalanced in 2020. A few years later, their forecast was confirmed by the estimation of an increase in overall patient demand of 40% over a thirteen year period (from 2012 to 2025) (Yang et al., 2014).

Despite the mentioned outlooks and new challenges, the oncology departments need to steadily maximize their service levels. To this end, the healthcare community looks for an improvement in service levels, which in turn impacts the patient quality of life. The reduction of the patient waiting time is considered one of the top priorities for patients in cancer departments (Gesell & Gregory, 2004). The main goal of the healthcare departments is to care for the highest number of patients in such a way as to reduce the patient waiting time and, simultaneously, increase patient satisfaction (Ahmed & Alkhamis, 2009). However, the oncology process involves diverse resources, both human and non-human, along with cooperation from pharmacies, all of which increases the complexity of the system. Therefore, simulation modelling was used to support decision-making policies. Simulation modelling also represents a risk-free and low budget method to assess the impact of potential changes on healthcare systems before implementing any intervention (Cassidy et al., 2019). In healthcare environments, Discrete Event Simulation (DES) is widely adopted for modelling and optimizing hospital workflows and other processes. According to this approach, variables and states change after a set of events occur at discrete time points and entities are simply data objects influencing the system's decision processes. In contrast, Agent-Based Simulation (ABS) modelling allows users to reproduce the actions and interactions of autonomous agents in order to handle the behaviour of a complex system. In healthcare contexts, people (*e.g.*, patients, doctors) can be represented by agents with an individual behaviour, but it is also possible to model rescue service vehicles and other resources using agents (Djanatliev & German, 2013).

Inspired by the studies performed on a real-life oncology department located in southern Italy, this paper presents a novel computer simulation model, which is configurable and adaptable to the needs of oncology departments cooperating with a detached pharmacy. The simulation model is based on a DES structure and was developed using Netlogo® (Wilensky, 1999), a programmable agent-based open-source platform that enables the realization of user-friendly intuitive interfaces. The presented model was designed to allow healthcare managers to recreate their oncology department in a virtual environment and easily test new configurations of the oncology process with the goal of reducing patient waiting time. The effectiveness of the proposed simulation model was verified through a case study based on an existing hospital oncology department. It is worth noting that, unlike similar configurations described in the literature, our model also considers the case in which the pharmacy department is detached from the oncology department and, therefore, therapies are gathered in batches by pharmacists and delivered through a courier. Once the proposed simulation model was validated, it was used to compare several 'what-if' configurations to identify better department configurations which minimize patient waiting times. The configurability and the free availability of the Netlogo agent-based framework, as well as the validation based on a real-life case study, represent the strengths of the proposed research. This paper provides several contributions to the scientific community:

i) It represents the first attempt to use an agent-based simulation model to investigate outpatient flow in a multi-stage oncology department where the pharmacy is detached from the department itself;

ii) It provides a configurable and adaptable tool that can easily be used by stakeholders to investigate alternative department configurations and also optimize the service level;

iii) A real-life situation is presented with the aim of testing and validating the effectiveness of the proposed approach;

iv) A series of findings arising from an ANOVA analysis allows the readers to assess how some organizational aspects may affect the performance of oncology departments.

The paper is organized as follows. After a comprehensive literature review, the proposed simulation model is introduced and described in detail. Then, the application to the case study is presented and the model is validated by comparing the behaviour of the real oncology department to the simulated one. Finally, a Design Of Experiments (DOE) is carried out with the aim of identifying more effective configurations of the oncology department under investigation. The best configuration in terms of patient waiting time reduction is identified and the managerial implications resulting from the present study are further discussed. Finally, the conclusion and future research directions are summarized.

## 2. Background and related work

Simulation tools are increasingly used in healthcare management, along with other Operational Research/Management Science (OR/MS) methods (for example the dynamic optimization in the work of Hahn-Goldberg et al., 2014 or the stochastic programming in the work of Demir et al., 2021). Often, computer simulation is employed to virtually evaluate 'what-if' configurations of health departments, so healthcare managers can assess the impact of potential changes on health systems without implementing them (Cassidy et al., 2019; Gunal, 2012; Salleh et al., 2017). In general, computer simulation can be classified as: Discrete Event Simulation (DES), System Dynamics (SD) or Agent-Based Simulation (ABS).

DES methodology deals with real systems which have a strong queue structure that can be modelled in discrete periods, where the process can be described stochastically. In this approach, patients are represented by 'entities' that go through different processes of the system. DES has been widely used to support healthcare decision-making. For example, Abo-Hamad and Arisha (2014) and Demir et al. (2017) merged the DES model with typical decision support tools (*e.g.*, balanced scorecard) in an emergency department, while Luo et al. (2018) applied DES in a radiology department to study how to reserve capacity for emergency and non-emergency patients.

On the other hand, SD is typically adopted to model health systems at an aggregate level. Introduced by Forrester (1958), it is based on differential equations and is used to capture the macro-level dynamics of a complex system under study. In this respect, Rashwan et al. (2015) modelled the flow of elderly patients to study the impact of various system parameters on the issue of acute bed blockage in the Irish healthcare system.

Edaibat et al. (2017) used SD simulations to assess the impact of health information exchange (HIE) adoption policies in hospitals located in the State of Maryland.

Finally, a considerable attention has being focused on ABS modelling in the OR context (Abar et al., 2017; Siebers et al., 2010) and for health systems as well (Cassidy et al. 2019; Gunal, 2012; Sulis et al., 2020). ABS uses 'agents' to model the flow of people (Metzner, 2019) as it allows writing specific instructions that control the behaviour and the interaction between agents in the system (Gunal, 2012; Mustafee et al., 2010). Several contributions reveal that ABS modelling is used to enhance the performance of healthcare departments. Yousefi and Ferreira (2017) combined ABS with decision-making techniques to re-allocate resources in an emergency department. Fragapane et al. (2019) developed an ABS model to enhance internal hospital logistics by examining the status of the goods' delivery system and evaluating potential improvements. Saeedian et al. (2019) and Ajmi et al. (2019) used the ABS approach to reduce indicators related to patients' pathways, such as total waiting time or length of stay, in surgery and emergency departments, respectively.

Another in the field of simulation concerns an innovative paradigm denoted as hybrid simulation. The hybrid simulation modelling consists of combining different types of simulation techniques (Brailsford et al., 2019). Recently, Olave-Rojas and Nickel (2021) developed a hybrid simulation framework combining ABS and DES to address the complex problem arising from Emergency Medical Services. Another research stream focuses on the integration between simulation and optimization related techniques. For example, Vahdat et al. (2019) used a simulation-optimization approach to study a paediatric orthopaedic outpatient clinic. Ordu et al. (2020) used a forecasting-simulation-optimization (FSO) strategy as a means of coping with the resource capacity needs of an entire hospital.

As far as oncology departments are concerned, Sepúlveda et al. (1999) and Baesler and Sepúlveda (2001) can be considered the pioneers of decision-making through simulation in oncology departments. Nowadays, these studies still represent a source of inspiration for researchers that aim to investigate the patient flow in oncology departments. Some research work implement simulations to examine the performance of the department by focusing primarily on the administration of chemotherapy treatment. Ahmed et al. (2011) employ simulation to propose new appointment scheduling rules with the aim of increasing both throughput (*i.e.*, number of patients per day) and treatment chair utilization. Baril et al. (2016a) and Baril et al. (2017) studied the nurses' tasks in an oncology department with the goal of reducing their workload. Baril et al. (2020) examined the workload of nursing staff in relation to the administration of patient treatment, considering both physical and mental workload. Other studies also included the activities of the pharmacy, which consists of preparing the therapies required by the oncology department (Alvarado et al., 2018; Arafeh et al., 2018; Baril et al., 2016b; Liang et al., 2015; Woodall et al., 2013) and delivering the therapies to the oncology department by a courier (Arafeh et al., 2018). Woodall et al. (2013) assessed the impact of nurses' unavailability on patient waiting times. Interestingly, Liang et al. (2015) proposed a robust DOE to support healthcare managers in the decision-making process by investigating the impact of various experimental factors on things such as the number of patients per day, number of chairs, etc. Baril et al. (2016b) combined simulations with a business game in a Kaizen event, *i.e.*, a workshop whose goal is to encourage the continuous improvement of a specific area or process (Farris et al., 2009). The authors compared a series of alternative management configurations and pointed out the need to include pharmacists in the Kaizen event. Alvarado et al. (2018) developed a simulation model to analyse operational strategies related to the management of patients' appointments in an oncology clinic.

In brief, this paper presents a computer simulation model to assess the patient flow of an oncology department, with the aim of identifying a series of actions that are capable of reducing the patient waiting time. To the best of our knowledge, this is the first time an agent-based simulation model has been developed to investigate oncology chemotherapy departments where agents reproduce patients, doctors, nurses and auxiliary resources. Although several agent-based packages are available, we deployed Netlogo® modelling software as it is considered a user-friendly tool that makes it possible for anyone to simulate complex physical systems (Cabrera et al., 2012; Chiacchio et al., 2014; Liu et al., 2017; Saeedian et al., 2019; Sulis et al. 2020; Taboada et al., 2011, 2012; Yousefi & Ferreira, 2017).

## 3. Model description

To describe the proposed model, the STRESS-ABS scheme according to the guidelines introduced by Monks et al. (2019), was used. This section is divided into two sub-sections: the former describes in detail the patient flow and the patient classification whilst the latter explains the dynamics of the proposed simulation model.

### 3.1 Problem formulation

In a generic day-hospital oncology department, the patients $p_k$ ($k = 1, ..., P$) are treated and discharged on the same day. The main resources involved in the care process are the oncologists, $o_j$ ($j = 1, ..., O$), the nurses, $n_l$ ($l = 1, ..., N$), and the therapy chairs, $c_i$ ($i = 1, ..., C$). Furthermore, each oncology department interacts with the pharmacy department, which in turns employs a number of pharmacy technicians, $d_z$ ($z =$ $1, \ldots, D$), for the therapy preparation process.

The oncology process can be regarded as a three-stage hybrid flow shop (Bouras et al., 2017; Hahn-Goldberg et al., 2014) with limited human resources, denoted in literature as *HFS/HR* problem (Costa et al., 2020). As depicted in Figure 1, the $k$-th patient, $p_k$, receives oncology services through the following three stages:

1) **Medical consultation:** Each patient arrives at the department and meets the nurse at reception for registration. Then, they are assigned to the $j$-th oncologist, $o_j(p_k)$, who defines the treatment protocol and assures the continuity of care of the patient. The treatment protocol specifies all the necessary information for the care path of the patient, such as therapies to be used for the treatment. Before starting the chemotherapy treatment, every patient has to meet the provided oncologist for a preliminary medical consultation. The duration of the medical consultation, $Tc(p_k)$, depends on the health status of the patient. The oncologist monitors the patient's health and evaluates the blood exams taken from the patient one or two days prior in the same hospital or in an external laboratory. Therefore, it is assumed that the results from the blood exams are available when the patient meets the oncologist. Finally, based on the patient's health conditions, the oncologist sends the type and dose of the therapy to the pharmacy;

2) **Therapy preparation:** When the pharmacy receives the request, the pharmacy technicians start the therapy preparation process, whose preparation time, $Tp(p_k)$, depends on the type of therapy. This process occurs after the medical consultation, rejecting any anticipatory therapy preparation policy, since, in case of absence or of unsatisfactory health status of the patient, the risk of wasting expensive therapies increases (Hesaraki et al., 2019). When the therapy is ready, it is delivered to the oncology department with a therapy delivering time, $Td$, which strictly depends on the location of the pharmacy and its distance from the oncology department. If the pharmacy is located far away from the oncology department, a courier service is necessary and therapies will be carried in batches, $b_w$ ($w = 1, \ldots, B$);

3) **Chemotherapy's administration:** Once the therapy arrives in the oncology department, the chemotherapy treatment of patient $p_k$ may start if both a nurse and a treatment chair are available. In this case, the setup task can be accomplished by a nurse, who prepares the patient for the chemotherapy treatment. Each nurse can prepare only one patient at a time, although during the treatment time of patients, $Tt(p_k)$, any nurse may simultaneously monitor the infusion process of $N_{max}$ patients, which in literature is usually set to four (Baesler & Sepúlveda, 2001; Baril et al., 2020). In addition, conforming to Demir et al. (2021), a nurse can monitor the infusion process of multiple patients while performing the setup process of a patient. Finally, when the therapy process is completed, the same nurse releases the patient who can leave the oncology department and go back home.

It is worth specifying that some patients do not need to undergo all the aforementioned processes. Patients can be classified into three categories depending on their daily pathway (Liang et al., 2015):

- **Standard patients**, or '*OC type*' patients, $p_{k_1}^{OC}$ (with $k_1 = 1, \ldots, P^{OC} | P^{OC} < P$), go through all the stage of the oncology department, as described above;

- **Repetitive patients**, or '*C type*' patients, $p_{k_2}^C$ (with $k_2 = 1, \ldots, P^C | P^C < P$), are allowed to skip the medical consultation, as they have already met the oncologist and received treatment the day before;
- **Control patients**, or '*O type*' patients, $p_{k_3}^O$ (with $k_3 = 1, \ldots, P^O | P^O < P$), do not need any therapy since they have successfully completed the provided chemotherapy protocol and they only require a periodical consultation with the oncologist.

It has to be specified that, in the same day, the oncologists can meet both $P^{OC}$ and $P^O$ patients. Since the continuity of care is usually assured for all patients, even $P^O$ patients have to be assigned to a specific oncologist, $o_j(p_{k_3}^O)$. The sum of the three categories of patients has to be equal to the total number of patients (*i.e.*, $P^{OC} + P^C + P^O = P$)

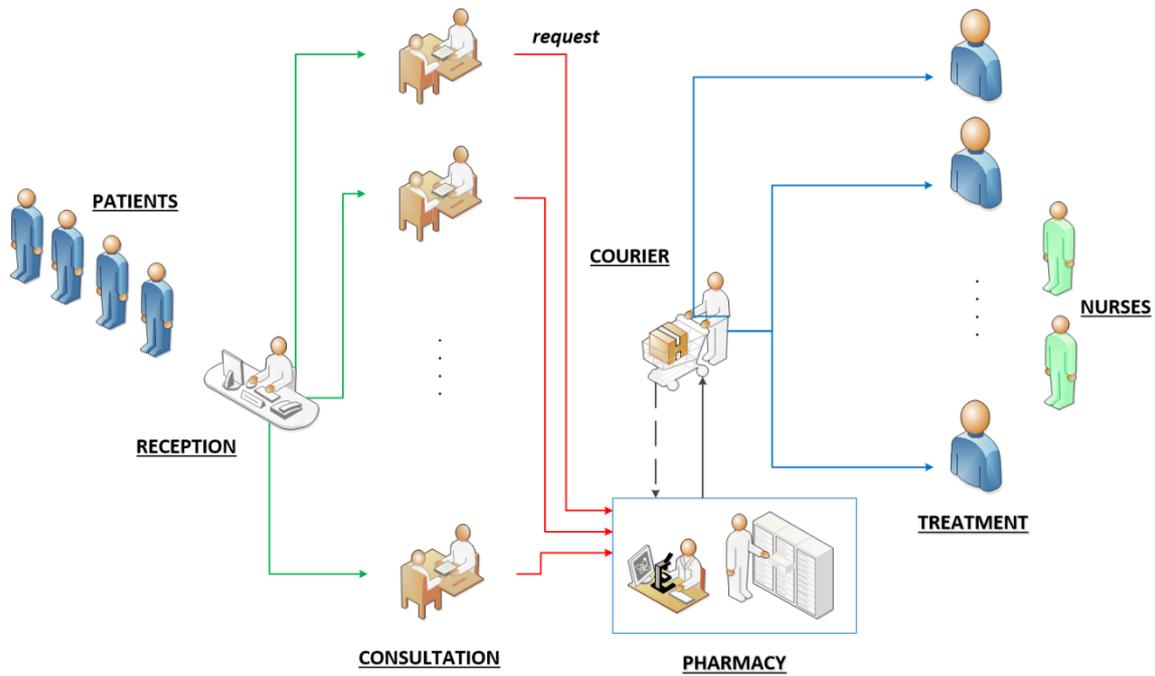

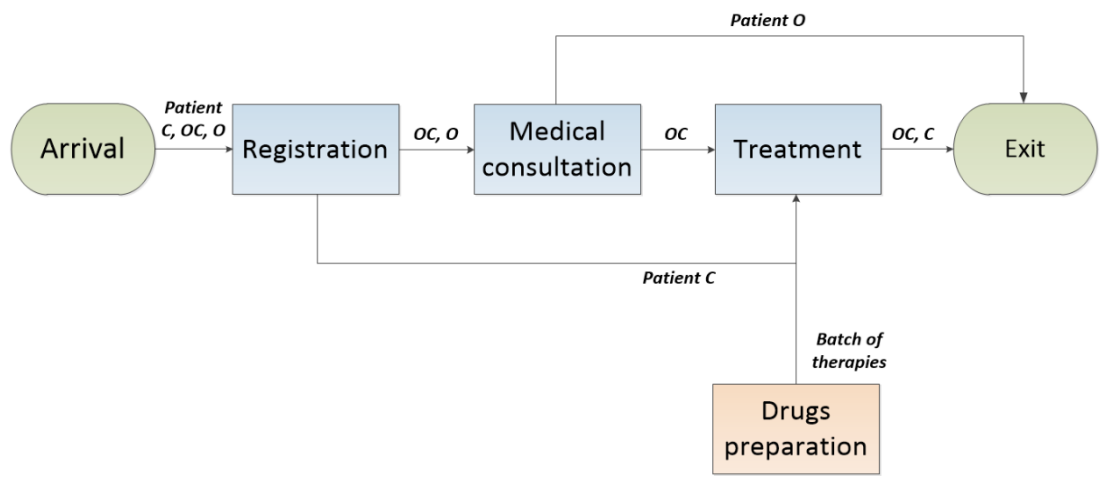

Figure 1. Description of the three-stage daily oncology care process. For each stage, we used different coloured arrows. The patient classification is also considered in the figure (Patient C = repetitive patients; patient OC = standard patients; patient O = control patients).

## 3.2 The simulation model

A healthcare setting can be seen as a complex system and the computer simulation represents a valuable tool to support the decision-making process. It allows users to identify the factors which influence the patient waiting time and possible bottlenecks in the systems under investigation. Figure 2 represents the graphical visualization of the proposed simulation model developed in the agent-based Netlogo® environment (Wilensky, 1999), including a key depicting the model agents. The simulation model is available from authors on reasonable request. The main features are described in the following sub-sections.

*3.2.1 Layout of the model*

A general layout of the model was defined to emulate the patient flow in the oncology departments. Considering that the patient waiting time does not depend on the location of the rooms in the department, there is no need to import the exact layout of an oncology department into the simulation model. To this end, two main assumptions can be considered in the model: *i)* the layout of the model is qualitative; *ii)* the time needed for each patient to relocate from one room to another is negligible. The layout of the model includes the following main rooms:

- The welcome room, where the patient meets the nurse at reception for registration;
- The first waiting room, where the patient waits for the medical consultation;
- The oncologists' room, where the patient meets their oncologist for the medical examination;
- The nurse room, where the courier delivers the batches of therapies;
- The second waiting room, where the patient waits for the treatment;
- The treatment room, where the patient undergoes the treatment monitored by the nurses.

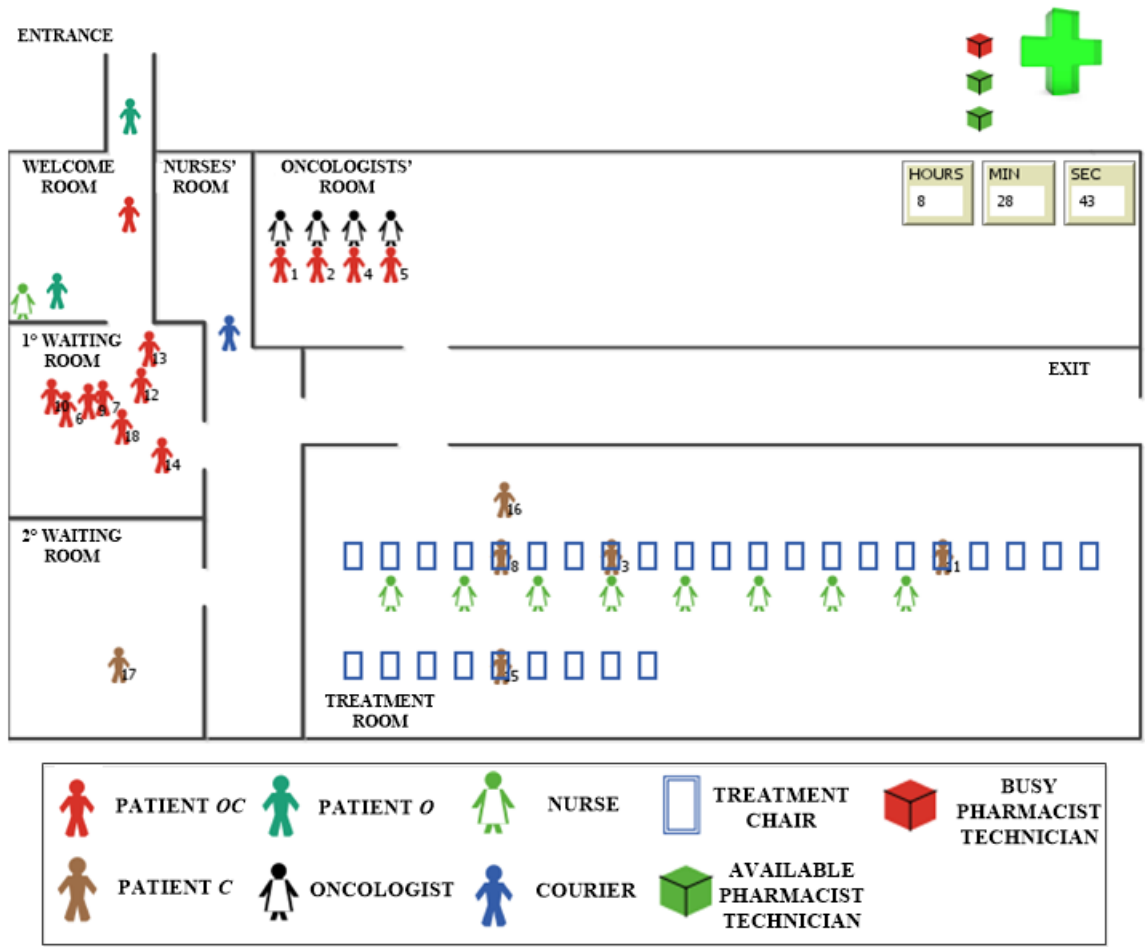

Figure 2. Agent-based framework of the simulation model.

The object located in the top-right corner of the simulation framework is the pharmacy. Finally, the simulation time clock (with a resolution of seconds) is visible on the top-right side of the agent-based simulation model.

*3.2.2 Modelling agents*

Each simulation run represents a single day in the oncology department, which starts at 08:00 AM and ends when all the treatments are concluded, which matches what happens in a real-life scenario. Patients and human resources are represented by two types of agents: *moving* agents, which move freely within the system, or *fixed* agents that occupy a specific location. Specifically, the patients and the courier act as moving agents, while the other resources play as fixed agents. For each simulation run, every patient agent is created in accordance with a vector of patients' arrival times, defined as *arrival_time_list*. The $k$-th patient, $p_k$, can move through the rooms previously described, following a path that depends on their classification, indicated by the agent's colour. Red agents are the standard patients, $P^{OC}$, the brown agents are the repetitive patients, $P^C$, and the green agents are the control patients, $P^O$. Each patient may interact with four types of resources: a nurse at reception, the oncologist for the medical consultation, the chair and the nurse for the treatment.

According to the problem formulation in Section 3.1, $P^{OC}$ patients follow the whole therapy pathway, $P^O$ patients are discharged after the medical consultation and $P^C$ patients are allowed to skip the medical consultation. All the patients start the medical consultation or the treatment based on the status of the resources involved in the related processes, which can be denoted as 'busy' or 'available'. In the case of the medical consultation, a patient $p_k$ is allowed to enter their oncologist's room, $o_j(p_k)$, only if the latter is available. The First In First Out (FIFO) policy is adopted to decide the order of patients for the oncologist visit. Finally, a patient $p_{k_1}^{OC}$ or $p_{k_2}^{C}$ goes to the treatment room if at least one chair, $c_i$, and one nurse, $n_l$, are 'available' and the courier has also delivered their therapy. As described in the model description (see Section 3.1), a nurse can setup only one patient at a time and can simultaneously monitor up to $N_{max}$ persons. In this regard, the nurses' agents are characterized by a *setup_status* and a *monitor_status* that can be 'busy' or 'available'. In fact, a patient starts the treatment if both *setup_status* and *monitor_status* of a nurse are simultaneously 'available'. A vector called *monitoring_patient_list* is created to record the patients monitored by the nurse. If the length of *monitoring_patient_list* is lower than provided limit, $N_{max}$, then the

*monitor_status* is 'available'.

With respect to pharmacy resources, each pharmacy technician, which is handled as a fixed agent, can prepare only one therapy at a time and pre-emption is not allowed. In Figure 2, pharmacists are represented by three boxes, whose colour indicates when each of them is available/unavailable (*i.e.*, green/red) to prepare therapies. The behaviour of agents related to the pharmacy strictly depends on the specific list of therapy requests coming from the oncologists, named *request_list*. If the list is empty, the agents are 'available' and the related box of the simulation framework becomes green. Otherwise, the agents status returns to 'busy' and the box becomes red. In this case, the therapy being prepared is registered in a vector called *wip_list*. When the preparation of a therapy is completed, another vector named *ready_list* is updated with the information of the therapies which need to be delivered. Once the length of the *ready_list* equals the provided batch size, the courier picks up the ready batch and delivers it from the pharmacy to the oncology department. At this point, a new vector denoted as *delivery_list* contains the information of the therapies that are being transported by the courier. Simultaneously, these therapies are removed from the *ready_list* and a new batch size is defined for the next therapies to be prepared and delivered.

As mentioned earlier, the courier for delivering the therapies is configured as a moving agent and is depicted in blue in Figure 2. It is assumed that the courier is exclusively tasked with carrying the therapy batches to the oncology department. The proposed simulation model also handles the round-trip of the courier from the oncology department to the pharmacy. The courier delivery time, $Td$, is an input variable which must be set by the analyst. Interestingly, if $Td$ is set to zero, it is assumed to model an in-house pharmacy. When the courier arrives to the oncology department, a specific

*therapy_flag* becomes 'true' to indicate that the patient's treatment may start.

*3.2.3 Communication between agents*

The simulation model is characterized by multiple interactions between agents. When communication exists between agents, one agent sends an input to another agent, causing an output, *i.e.*, a certain behaviour of the latter agent. The model includes three types of communication (Yousefi & Ferreira, 2017): *i)* one-to-one; *ii)* one-to-n; *iii)* one-to-location. One-to-one communication happens when a single agent interacts with another agent, as in the case of the interaction between a patient and an oncologist. In this case, the arrival of the patient in the oncologist's room (input) changes the status of the oncologist to 'busy' (output). One-to-n communication occurs when a single agent communicates with a group of agents (for example, the communication between the courier arriving in the department and the group of nurses to notify that the batch of therapies was delivered). Finally, one-to-location communication exists when an agent communicates with agents in a different location, such as when an oncologist communicates with the technicians at the pharmacy in order to request the preparation of the patient's treatment. Table 1 shows the kinds of communications involved in the simulation model for oncology departments.

| Input agent | Output agent | Type of communication | Description |
| --- | --- | --- | --- |
| Nurse at reception | Patient | One-to-n | The nurse at reception registers patients according to a FIFO rule. |
| Oncologist | Patient | One-to-n | Each oncologist receives a patient on the basis of the FIFO rule. |
| Patient | Oncologist | One-to-one | The arrival of the patient in the oncologist's room makes the assigned oncologist busy. |
| Oncologist | Pharmacy | One-to-location | The oncologist sends a request for a new treatment preparation to the pharmacy. |
| Pharmacy technician | Courier | One-to-one | A pharmacy technician notifies the courier that the batch is ready to be delivered. |
| Courier | Nurse | One-to-n | The courier arrives at the department and notifies the group of nurses that the therapies have been delivered. |
| Nurse | Patient | One-to-n | Nurses allow patients to start the treatment once their therapy is at the department and a chair is available. Again, the first patient arrived at the waiting room is the first served. |
| Patient | Nurse | One-to-one | The arrival of the patient in the treatment's room makes the nurse 'busy'. |

Table 1. Communication between agents in the proposed simulation model.

## 4. Case study

The proposed simulation model was applied to improve the quality of services provided by a real-world oncology department located in southern Italy. The goal of the project was to analyse the performance of the oncology department in its current configuration and, subsequently, to find new configurations capable of reducing the patient waiting times. The preliminary phases of the project were the following. First, briefings with the clinic's employees were held to define: *i*) the features of the oncology department; *ii*) the key performance indicators. Over a three week period, the project team, which includes clinicians, members of the oncology department and developers of the simulation model, performed an intense time study on the tasks related to the different oncology processes described earlier. Once the data had been collected, a statistical analysis was performed with the aim of finding the stochastic distributions of the main input variables of the simulation model.

*4.1 Key Performance Indicators (KPIs)*

It is well known that cancer diseases dramatically affect the physical and emotional status of suffering individuals. In this context, reducing the patient waiting time is the main objective so as to enhance the quality of cancer treatment within facilities (Gesell & Gregory, 2004), which is recognized as the primary source of patient dissatisfaction (Aboumatar et al., 2008; Edwards et al., 2017; Gourdji et al., 2003). In light of the previous considerations, in this paper the total flowtime, *F* (*i.e.,* the sum of the *length of stay* of patients), was adopted as a key performance indicator (KPI). The length of stay consists of the total time a patient spends in the oncology department, *i.e.*, the time interval ranging from the time he/she is registered at reception to the end of the chemotherapy treatment. Particularly, the mean flow time, from now on denoted as $\bar{F}$, was selected to measure the performance of any department configuration in the successive analyses. Furthermore, two additional indicators were engaged to compare the *status-quo* of the oncology department with the simulated configurations, namely the mean patient waiting time, $\overline{WT}$, and the system efficiency, $Eff$, calculated as follows:

$$Eff = \frac{\bar{F} - \overline{WT}}{\bar{F}} \cdot 100 \qquad (1)$$

*4.2 Data collection and statistical distributions*

A time study covering three working weeks was carried out to collect the experimental data related to the department status-quo. During that period, the healthcare department, which consisted of 3 oncologists, 13 chairs, 1 nurse at reception and 3 nurses for the treatment, received 28 patients on average per day. Four patients who are receiving treatment can be simultaneously monitored by one nurse. A single pharmacy technician working in an external pharmacy is dedicated to the preparation of the oncology therapies. A single auxiliary courier is employed to deliver the therapy batches from the pharmacy to the oncology department.

Table 2 reports model parameters and stochastic distributions obtained by analysing the aforementioned status-quo related data. The number of patients per day is derived from a *normal* distribution with mean 28.07 and standard deviation 3.94. As stated above, usually these patients undergo two different processing stages: medical consultation and chemotherapy administration, which starts after the therapy delivery. Among the patients, 22.32% need only the medical consultation ($P^O$), while 6.18% attend only the chemotherapy's administration monitored by the nurse ($P^C$), and the remaining 71.50% are classified as standard patients ($P^{OC}$). The experimental analysis revealed that the arrival times for each type of patient can be handled by considering five time windows, each one related to a different occurrence probability. Therefore, once the time interval is selected, every patient arrival time is drawn from an *uniform* distribution U[0, 59] in minutes. For $P^O$ and $P^{OC}$, the oncologist is assigned to the patient using a random criterion as soon as the patient agent is created. The duration of the medical consultation is derived from a *uniform* distribution U[5,35], in minutes. This uniform distribution is adopted for both $P^{OC}$ and $P^O$ patients. The order of patients for the medical consultation is decided using the FIFO policy. Regarding the therapy's preparation, they can be classified into three typologies based on preparation time (short, medium and long. Therapies are delivered in batch sizes, which may vary between 2 and 12, depending on courier availability and pharmacy workload. A batch may contain any type of therapy, while the batch size may vary at every courier pick up. The courier takes 10 minutes to deliver the therapies to the department and another 10 minutes to return to the pharmacy.

However, there is a 26.53% probability of delay due to traffic congestion in each direction. Finally, the experimental studies conducted on the department showed that the treatments can be classified into five types, each one involving a different time duration. Notably, each treatment can be executed according to a specific occurrence probability and its duration implies the setup time. It is assumed that there is no relationship between the treatment duration and the therapy's preparation time. The time needed by a nurse to release a patient after the treatment can be considered negligible (Hesaraki et al., 2019).

| Descriptors of process | Values or probability distribution |
|---|---|
| **Patients** | |
| Number of patients ($P$) | N(28.07,3.94) |
| Classification of patient | |
|     Standard patient ($P^{OC}$) | 71.50% |
|     Repetitive patient ($P^C$) | 6.18% |
|     Control patient ($P^O$) | 22.32% |
| **Arrival time** | |
| 08:30-09:30 | 56.58% |
| 09:31-10:30 | 12.54% |
| 10:31-11:30 | 5.81% |
| 11:31-12:30 | 13.76% |
| 12:31-13:30 | 11.31% |
| **Registration** | |
| Number of nurses at reception ($NR$) | 1 |
| Duration (min) of registration | 1 |
| **Medical consultation** | |
| Number of oncologists ($O$) | 3 |
| Duration (min) of medical consultation ($Tc(p_k)$) | U(5,35) |
| Assignment of patient-oncologist ($o_j(p_k)$) | Random |
| **Pharmacy** | |
| Number of pharmacy technician ($D$) | 1 |
| Duration (min) of therapy's preparation ($Tp(p_k)$) | |
|     Short preparation | U(1,5) |
|     Medium preparation | U(6,10) |
|     Long preparation | U(11,27) |
| Probability of typology of therapy's preparation | |
|     Short preparation | 71.38% |
|     Medium preparation | 20.34% |
|     Long preparation | 8.28% |
| **Therapies' delivery** | |
| Number of couriers ($\alpha$) | 1 |
| Batch size | U(2,12) |
| Duration (min) of delivery ($Td$) | |
|     Delivery without delay | 10 |
|     Delivery with delay | 10 + U(2,10) |
| Probability of delay in delivery | |
|     Delivery without delay | 73.47% |
|     Delivery with delay | 26.53% |
| **Treatment administration** | |
| Number of chairs ($C$) | 13 |
| Number of nurses ($N$) | 3 |
| Treatment duration (min) ($Tt(p_k)$) | |
|     Type 1 | U(15,60) |
|     Type 2 | U(61,120) |
|     Type 3 | U(121,180) |
|     Type 4 | U(181,240) |
|     Type 5 | U(241,300) |
| Probability of treatment occurrence | |
|     Type 1 | 30.13% |
|     Type 2 | 38.91% |
|     Type 3 | 14.23% |
|     Type 4 | 12.13% |
|     Type 5 | 4.60% |

Table 2. Model descriptors.

## 5. Experimental results

The verification and validation process were performed to verify if the simulation model was consistent with the problem description and if the outcomes of the simulations reproduced the *status quo* of a typical day in the oncology department. Then, a DOE was arranged to use the validated simulation model to test different department configurations and improve the performance of the department using the validated simulation model. In light of the multitude of stochastic parameters, a stochastic simulation approach was adopted for all the numerical investigations to assure the robustness of the proposed analysis. Therefore, each KPI was evaluated in terms of its expected value:

$$E(KPI) = \frac{\sum_{\omega=1}^{\Omega} KPI(\omega)}{\Omega} \qquad (2)$$

where $\omega$ is the replicate of a certain department configuration and $\Omega$ is the whole set of replicates.

### *5.1 Verification and validation of the simulation model*

A preliminary step of any simulation model consists in demonstrating that it provides credible results (Balci, 2003; Roza et al., 2017). To this end, Verification and Validation (V&V) techniques are generally carried out to assure the effectiveness of a simulation model (Kleijnen, 1995). Specifically, the verification process assures that the conceptual model of the problem was transformed into a computer simulation model with sufficient accuracy (Robinson, 1997). The well-structured debug tool of NetLogo® and its model visualization were used to perform a dynamic verification test of the simulation model, which is widely used in literature (Sargent, 2013). Validation is necessary to demonstrate the efficacy of the model in reproducing the actual performance of the system under investigation with a satisfactory approximation. Sargent (2013) classified several validation techniques that can be applied to a given simulation model. In this paper we adopted the '*Historical data validation*' technique, which compares the key performance indicators obtained by the presented simulation model with one obtained by analysing the *status-quo* related configuration, as shown in Table 3. Looking at the numerical outcomes, the actual performance of the oncology department in terms of the aforementioned KPIs are as follows:

- the mean flowtime, $\bar{F}$, is equal to 265.46 minutes, with a 95% confidence interval (CI) equal to [243.00; 287.92] ;

- the mean patient waiting time, $\overline{WT}$, is equal to 138.28 minutes, with a 95% CI equal to [123.14; 153.42];

- the efficiency, $Eff$, is equal to 47.97%, with a 95% CI equal to [45.41%; 50.53%];

For both the real and the simulation configurations, Table 3 reports the expected KPIs, the confidence intervals at 95% and the percentage deviation ($Dev$). $Dev$ is calculated as follows:

$$Dev = \left| \frac{E(KPI_{sim}) - KPI_{real}}{KPI_{real}} \cdot 100 \right| \qquad (3)$$

where $E(KPI_{sim})$ is the expected KPI resulting from the simulation model, while $KPI_{real}$ is the KPI's value of the *status-quo* of the oncology department. Interestingly, the percentage deviation values ($Dev$) reported in Table 3 confirm the validity of the developed simulative procedure. To further strengthen this outcome, the last column of the table reports the *p*-values resulting from the paired t-tests carried out for each KPI. The paired t-tests are used in order to assess if there exists any statistically significant difference between the means of the real and simulated configurations; *p*-values greater than 0.05 for each test pointed towards the effectiveness of the proposed simulation model in simulating the dynamics of the oncology department under investigation.

| KPIs | Real | 95% CI | Simulated (E(KPI)) | 95% CI | Dev | p-value |
|---|---|---|---|---|---|---|
| $\bar{F}$ Mean Flowtime [min] | 265.46 | (243.00;287.92) | 259.50 | (242.30;276.70) | 2.25% | 0.684 |
| $\overline{WT}$ Mean Waiting Time [min] | 138.28 | (123.14;153.42) | 133.99 | (116.54;151.44) | 3.10% | 0.730 |
| $Eff$ Efficiency | 47.91% | (45.41;50.53) | 48.84% | (45.21;52.47) | 1.94% | 0.742 |

Table 3. Validation of the simulation model with historical data (results from 15 working days of measurements).

## 5.2 Design of Experiments (DOE)

In order to explore alternative configurations of the oncology department, a full-factorial DOE was developed. DOE is a statistical method which enables the identification of the impact of a series of experimental factors on a response variable. The experimental factors, shown in Table 4, were suggested by the medical staff and were taken into consideration since the costs of implementation were low or negligible. Briefly, such factors can be described as follows:

1) **The number of couriers (α)**. This refers to the number of couriers employed to deliver the batches of therapies to the oncology department. Since only one resource is currently available for this task (level *A* in Table 4), the aim is to evaluate how an additional resource (level *B*) would affect the patient waiting time;

2) **The batch size (β)**. The second factor consists of the number of therapies that can be collected in a batch. Currently, the batch size is not fixed and can vary from two to twelve therapies. The objective is to assess if a fixed batch size can enhance the adopted KPIs and, at the same time, to evaluate if a smaller batch size is better than a larger one. To this end, three levels were considered: (*A*) fixed batch sized with three therapies; ($B$) fixed batch size with six therapies; ($C$) variable batch size (*i.e.*, corresponding to the current configuration);

3) **The appointment distribution ($\gamma$)**. The first level ($A$) provides three time-windows of one hour and thirty minutes, each one with the same probability of occurrence equal to 33%. Similarly, the second level ($B$) consists of five time-windows of one hour, each with a probability of 20%. Level $C$ describes the current case in which patients arrive at the oncology department conforming to five time-windows characterized by different occurrence probability (see Table 2);

4) **The daily number of patients ($\delta$)**. The last factor represents the average number of patients for each working day. Currently, every day the department takes care of about 28 patients (level $A$). The goal is to analyse how the performance changes when considering a higher number of patients. To this end, an additional level ($B$) with 31 individuals is considered, which corresponds to an increase of about 10% of patients per day. It is worth specifying that both levels refer to the mean of the *normal* distribution related to the number of patients per day (see Table 2) , while the standard deviation is kept constant at 3.94.

Notably, the current configuration of the oncology department is {A-C-C-A}, considering a one-to-one correspondence with the set of experimental factors {$\alpha$-$\beta$-$\gamma$-$\delta$}, respectively. We defined a full-factorial DOE, which involves $3^2 \cdot 2^2 = 36$ different configurations of the oncology department, in order to study the influence of the experimental factors on the performance of the department. In addition, to make the statistical analysis robust enough, $\Omega = 5{,}000$ different replicates at varying random seeds, each one simulating a different working day, were executed, thus achieving a number of $5{,}000 \cdot 36 = 180{,}000$ experiments. The DOE was performed on five virtual machines installed on a workstation equipped with an INTEL i9-9900 3.6 GHz 10 core CPU, 32Gb DDR4 2,666MHz RAM and Win 10 PRO OS. Since the computational time required to simulate each configuration is equal to about 5 seconds, approximately two days were needed to accomplish the whole DOE. Only the expected mean flowtime, $E(\bar{F})$, was used as a KPI, since the expected mean waiting time, $E(\overline{WT})$, and the expected efficiency, $E(Eff)$, are strictly related to the former. However, all KPIs will be used in the next analysis to stress the difference between the best configuration and the status quo.

| Factors | | Levels | | |
|---|---|---|---|---|
| Symbols | Description | A | B | C |
| $\alpha$ | Number of couriers | 1 | 2 | - |
| $\beta$ | Batch size | 3 | 6 | U(2,12) |
| $\gamma$ | Appointment distribution | 3 | 5 | 5* |
| $\delta$ | Capacity of the department | 28 | 31 | - |

(*) time intervals with different occurrence probabilities for the status quo configuration (see Table 2).

Table 4. Factors/levels involved in the design of experiments.

### *5.3 Analysis of results and managerial implications*

The analysis of variance (ANOVA) determines whether the experimental factors statistically influence the key performance indicators. To this end, an ANOVA analysis at 95% level of confidence was carried out in the Minitab® 2017 commercial package to evaluate the statistical significance of each factor. The numerical outputs from the ANOVA (see Table 5) show the results concerning the main effects. The plots related to the main effects are reported in Figure 3. The 2-way interactions are not reported in the table (but are available upon request) since no relevant findings were detected. Looking at the condensed ANOVA table, it is worth pointing out that the adjusted R-squared (*i.e.*, the adjusted coefficient of determination) is larger than 95%. A higher value of the R-squared demonstrates that the model fits the data of the analysis, thus confirming the robustness and the consistency of the proposed approach.

| Source | DF | F-value | p-value |
|--------|----|---------|---------|
| Model  | 19 | 40299.03 | 0.000 |
| $\alpha$ | 1 | 42.99 | 0.000 |
| $\beta$ | 2 | 292158.50 | 0.000 |
| $\gamma$ | 2 | 52500.10 | 0.000 |
| $\delta$ | 1 | 73713.83 | 0.000 |
|        |    |         | *Adjusted $R^2$ > 95%* |

Table 5. ANOVA Table.

With regards to the experimental factors, the $p$-value below 0.05 implies that they are statistically significant for the expected mean flowtime, $E(\bar{F})$, at 95% confidence level. The significance of the influencing factors on the mean flowtime is further exacerbated by related $F$-values. Indeed, the most important factors are usually identified by an $F$-value larger than 50 (Yu et al., 2018). The $F$-value associated with factor $\alpha$ reveals that the number of couriers may have a weak effect on the performance of the system, as confirmed by the related main effect plots in Figure 3. To this end, a paired $t$-test at 95% confidence was performed and confirmed that the null hypothesis assumption that the mean difference between the paired samples is zero (*i.e.*, $H_0: \mu_d = 0$) can be rejected. In conclusion, the mean flow time is statistically insensitive to factor $\alpha$.

Interestingly, Figure 3 related to factor $\beta$ shows that fixing the batch size at the lowest value (level $A$) would favour the mean flow time reduction, while the current strategy based on a random batch size (level $C$) negatively biases the mean waiting time of patients. As for $\gamma$, rendering the arrival of the patients smoother by introducing new appointment distribution strategies (*e.g.*, levels $A$ and $B$) makes the service level better than the actual one (level $C$). In particular, the strategy corresponding to level $B$ reduces patient waiting time by approximately 20 minutes on average. Finally, as for factor $\delta$, an increase in the number of patients (level *B*) slightly increases the patient waiting time but, it can be adequately compensated for by a larger number of patients that can be accepted daily without worsening the current performance of the oncology department.

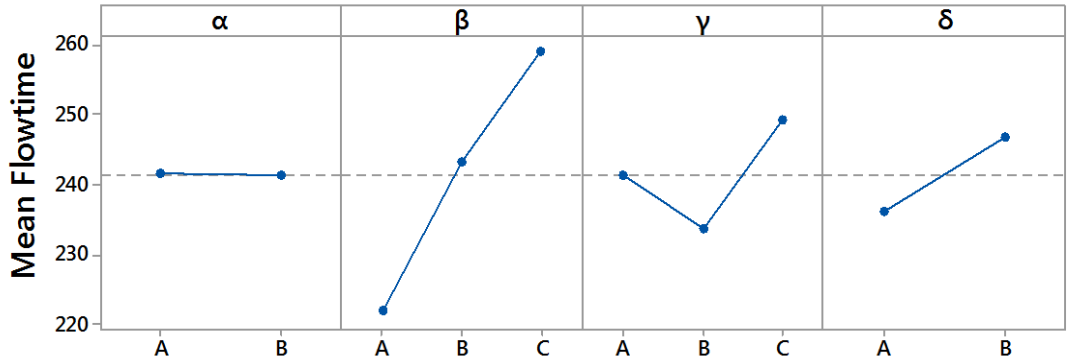

Figure 3. Main Effect Plots.

Table 6 shows the expected mean flowtime, $E(\bar{F})$, over the $\Omega = 5{,}000$ simulation replicates, performed for each combination of experimental factors. Notably, the performance of the *status quo* configuration is illustrated in the first row of the table, while the other configurations were sorted in ascending order of expected mean flowtime. Also, the confidence intervals of the expected mean flowtime for each configuration are reported in the last column. Looking at the table, configuration number 21 indicates the '*best configuration*' characterized by factors {B-A-B-A} and an expected mean flowtime $E(\overline{F_{21}})$ equal to 208.53. It is worth noting that the *status quo* configuration is one of the worst in terms of expected mean flowtime, along with the last four configurations in which the *β* factor always is set to the *C* level.

To sum up, the following managerial implications would arise from the proposed numerical analysis:

1) On the daily basis, the oncology department could save 40 minutes of patient waiting time by moving from a random batch size to a fixed batch size of three therapies. This improvement could be realized without investing additional funds;

2) Focusing on the patients' appointments could also reduce the patient waiting time. A uniform distribution of patients' arrival times through five time-windows of one hour emerges as a valid alternative to enhance the performance of the department without investing additional funds;

3) Looking at the best configuration, an increase in the number of patients per day (configuration number 22) would involve a slight increase in expected patient waiting time to about ten minutes on average. However, this configuration remains more successful than the status-quo configuration in terms of patient waiting time;

4) Since the number of couriers does not influence the expected mean flowtime, there would be no benefit from the addition of new resources dedicated to therapy delivery.

| Configuration No. | Factors | | | | $E(\bar{F})$ [minutes] | 95% CI [minutes] |
|---|---|---|---|---|---|---|
| | α | β | γ | δ | | |
| *17 (status-quo)* | *A* | *C* | *C* | *A* | *260.17* | *(259.22;261.12)* |
| 21 | B | A | B | A | 208.53 | (207.83;209.23) |
| 3 | A | A | B | A | 209.05 | (208.34;209.75) |
| 19 | B | A | A | A | 216.07 | (215.34;216.80) |
| 1 | A | A | A | A | 216.61 | (215.88;217.34) |
| 22 | B | A | B | B | 218.16 | (217.42;218.91) |
| 4 | A | A | B | B | 218.71 | (217.97;219.46) |
| 23 | B | A | C | A | 223.29 | (222.41;224.17) |
| 5 | A | A | C | A | 223.81 | (222.93;224.69) |
| 20 | B | A | A | B | 226.71 | (225.94;227.48) |
| 2 | A | A | A | B | 227.27 | (226.50;228.04) |
| 27 | B | B | B | A | 231.02 | (230.30;231.74) |
| 9 | A | B | B | A | 231.13 | (230.41;231.84) |
| 24 | B | A | C | B | 237.48 | (236.53;238.43) |
| 6 | A | A | C | B | 238.05 | (237.11;239.00) |
| 25 | B | B | A | A | 238.34 | (237.61;239.08) |
| 7 | A | B | A | A | 238.46 | (237.73;239.20) |
| 28 | B | B | B | B | 240.09 | (239.35;240.83) |
| 10 | A | B | B | B | 240.20 | (239.46;240.94) |
| 29 | B | B | C | A | 244.69 | (243.80;245.58) |
| 11 | A | B | C | A | 244.78 | (243.89;245.67) |
| 33 | B | C | B | A | 247.50 | (246.67;248.33) |
| 15 | A | C | B | A | 247.64 | (246.81;248.47) |
| 26 | B | B | A | B | 248.45 | (247.68;249.22) |
| 8 | A | B | A | B | 248.56 | (247.79;249.34) |
| 31 | B | C | A | A | 253.99 | (253.14;254.84) |
| 13 | A | C | A | A | 254.13 | (253.28;254.98) |
| 34 | B | C | B | B | 255.93 | (255.07;256.78) |
| 16 | A | C | B | B | 256.05 | (255.20;256.91) |
| 30 | B | B | C | B | 257.66 | (256.70;258.62) |
| 12 | A | B | C | B | 257.76 | (256.80;258.72) |
| 35 | B | C | C | A | 260.17 | (259.21;261.12) |
| 32 | B | C | A | B | 263.63 | (262.74;264.52) |
| 14 | A | C | A | B | 263.76 | (262.88;264.65) |
| 36 | B | C | C | B | 272.56 | (271.54;273.59) |
| 18 | A | C | C | B | 272.69 | (271.67;273.71) |

Table 6. Results of expected mean flowtime from the experimental campaign (5,000 replicates).

Finally, Table 7 compares the best simulated configuration and the simulated *status quo* in terms of expected mean flowtime, $E(\bar{F})$, expected mean waiting time, $E(\overline{WT})$, and expected efficiency, $E(Eff)$. Notably, Table 7 also reports the related 95% confidence intervals (CI) and the percentage deviations ($Dev$). The percentage deviations reveal that the best configuration reduces the expected flowtime, $E(\bar{F})$, by 19.85%, the expected mean waiting time, $E(\overline{WT})$, by 37.73% and increases the expected efficiency, $E(Eff)$, by 24.77%.

| KPIs | Simul. status quo | 95% CI | Simul. best config. | 95% CI | *Dev* |
|---|---|---|---|---|---|
| $E(\bar{F})$ | 260.17 min | (259.22;261.12) | 208.53 min | (207.83;209.23) | 19.85% |
| $E(\overline{WT})$ | 136.87 min | (136.01;137.73) | 85.23 min | (84.66;85.80) | 37.73% |
| $E(Eff)$ | 47.39% | (47.21;47.57) | 59.13% | (58.96;59.30) | 24.77% |

Table 7. Simulation results from 5,000 replicates/runs: status quo vs best configuration.

## 6. Conclusion

In this study, we developed a computer agent-based simulation model explicitly designed to be configurable and adaptable to the needs of oncology departments. The case in which the pharmacy is detached from the oncology department and, therefore, a courier service delivers batches of therapies is considered. The validity of the proposed model was demonstrated through a statistical analysis based on a set of experimental data obtained by studying an oncology department located in southern Italy. Consequently, a series of alternative configurations were tested through a robust simulation campaign based on a full-factorial design of experiments. The results were evaluated through an ANOVA analysis, revealing that a fixed batch size with a low number of therapies and an effective appointment strategy significantly decrease the patient waiting time. The best simulated configuration was selected and compared with the status quo by means of three main key performance indicators. This comparison shows that the expected patient waiting time can be reduced by 37.7% in percentage deviation and the expected department efficiency can be increased by 24.8%. Finally, the simulation model was delivered to the healthcare managers who implemented the "best configuration" to the chemotherapy department at hand. Due to the COVID pandemic we are not allowed to visit the department, but we received positive feedback from the medical staff regarding a significant improvement in service level. Hopefully, further data will be collected in the near future to support the validity of the proposed research.

The present paper reports an experimental campaign specifically conducted for the oncology department at hand. In fact, the proposed DOE was defined jointly with the healthcare staff of the hospital, with the aim of identifying an improved service configuration, without investing additional funds (according to lean principles). However, in accordance with the managing staff, further efforts will be dedicated to a future project to assess the impact of additional resources (*e.g.*, the number of pharmacy technicians or treatment chairs) and different queuing mechanisms on the same or new performance indicators (*e.g.*, overtime after the closing time or nurse workload). To this end, future research can be oriented towards either simulation-optimization approaches or hybrid simulation models, capable of adequately capturing macro- and micro-level dynamics of such complex healthcare systems.